\documentclass[aps,reprint,nofootinbib,preprintnumbers,twocolumn,floatfix,superscriptaddress,prl,showpacs]{revtex4-1}
\usepackage{physics}
\usepackage{natbib}
\usepackage{amsmath,amssymb}
\usepackage{graphicx}
\usepackage[dvipsnames]{xcolor}
\usepackage[caption=false]{subfig}
\usepackage{float}
\usepackage[pdftex]{hyperref} 

\newcommand{\avg}[1]{\langle #1 \rangle} 

\begin{document}
	\title{Super sensitivity and super resolution with quantum teleportation}
	\author{J. Borregaard}
	\affiliation{QMATH, Department of Mathematical Sciences, University of Copenhagen, Universitetsparken 5, 2100 Copenhagen, Denmark}
	\author{T. Gehring}
	\author{ J. S. Neergaard-Nielsen}
	\author{U. L. Andersen}
		\affiliation{Center for Macroscopic Quantum States (bigQ), Department of Physics, Technical University of Denmark, Fysikvej 5, 2800 Kgs. Lyngby, Denmark}
	\date{\today}
	
	\begin{abstract}
We propose a method for quantum enhanced phase estimation based on continuous variable (CV) quantum teleportation. The phase shift probed by a coherent state can be enhanced by repeatedly teleporting the state back to interact with the phase shift again using a supply of two-mode squeezed vacuum states. In this way, both super resolution and super sensitivity can be obtained due to the coherent addition of the phase shift. The protocol enables Heisenberg limited sensitivity and super-resolution given sufficiently strong squeezing. The proposed method could be implemented with current or near-term technology of CV teleportation.      
	\end{abstract}
	\maketitle
	
Quantum correlations can be used in a number of ways to enhance metrological performance~\cite{giovannetti2004quantum,giovannetti2011advances,Escher2011,Deganreview2017}. Highly entangled states such as NOON and GHZ states can enable Heisenberg limited sensitivity yielding a square root improvement with the number of photons over the standard quantum limit (SQL)~\cite{pan2012,Rafal2015,Bouchard2017}. This kind of improvement is particularly useful for probing fragile systems where photon damage limits the allowed number of probe photons. This can be the case in, e.g. imaging of biological systems such as live cells~\cite{Frigault2009}, molecules~\cite{Celebrano2011}, and proteins~\cite{Piliarik2014}. While this effect has been demonstrated in experiments for small probe sizes~\cite{Mitchell2004,Afek2010,Wolfgramm2012,Facon2016}, scaling up the size of the entangled states remains a technological barrier due to their fragility to loss and noise. Other strategies based on the more experimentally accessible squeezed vacuum states have also shown to beat the SQL in various settings~\cite{Caves1981,Ligo2011,Ligo2013,Taylor2013,Berni2015,Schafermeier2016}. An alternative strategy is to perform multi-pass protocols with a single probe. This enables both Heisenberg limited sensitivity and super-resolution~\cite{note1} for phase estimation without entangled resources by applying the phase shift to the same probe multiple times~\cite{giovannetti2006quantum,vanDam2007,Demkowicz2010}. Its experimental demonstration was realized by surrounding the phase shift system with mirrors to measure a transversally distributed phase shift~\cite{Higgins2007} or an image~\cite{Juffmann2016} with Heisenberg-limited sensitivity. While these approaches have demonstrated the effect of sub-shot noise scaling without entanglement, the former demonstration could only measure a transversally distributed phase shift while both demonstrations were based on post-selection, rendering the efficiency very low. 

\begin{figure} [t] 
\centering
\includegraphics[width=0.49\textwidth]{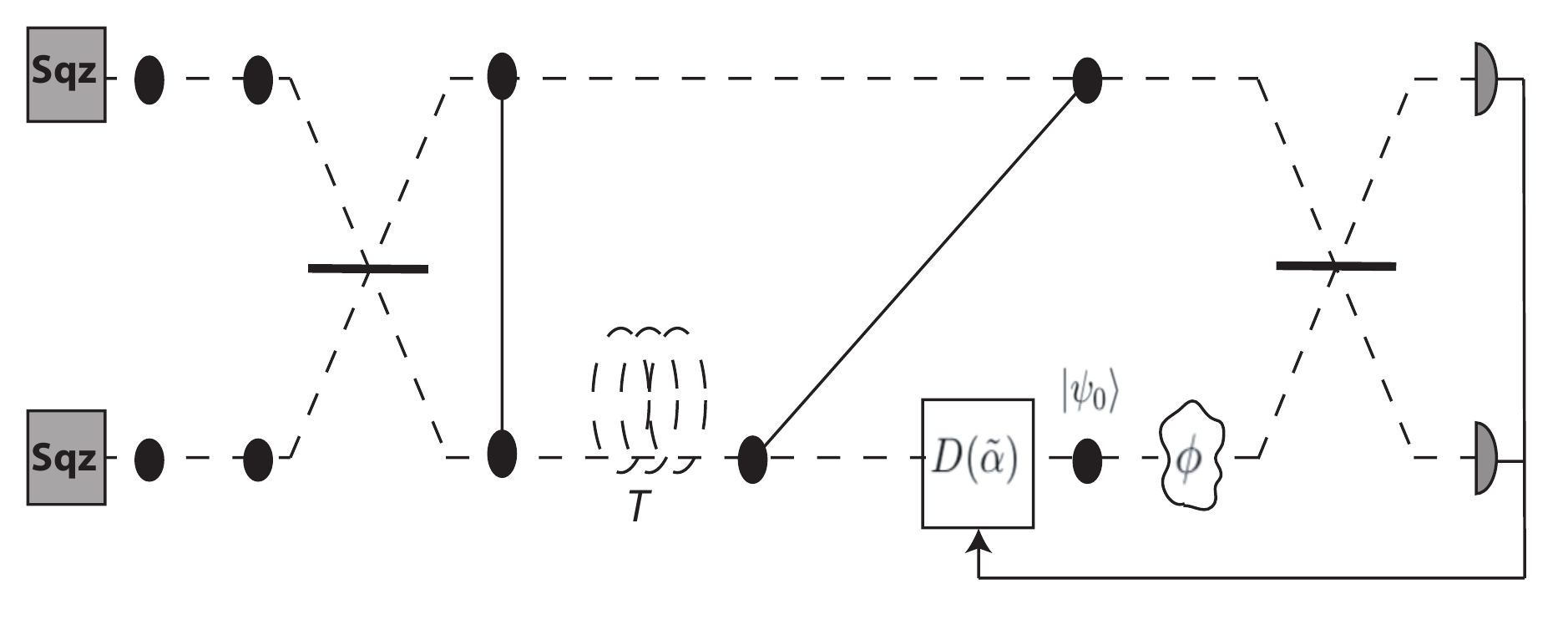}
\caption{Sketch of the setup considered. Consecutive pulses of two-mode squeezed vacuum states (illustrated by connected dots) are being supplied by interferring the outputs of two single mode squeezers on a balanced beam splitter. The second mode is delayed by some time $T$ such that it coincides in time with the first mode of the next pair. The second mode is subject to feedback (quadrature displacement) based on previous measurements before a phase shift $U(\phi)=e^{i\phi\hat{n}}$ acts on it. Here, $\hat{n}$ is the photon number operator. Initially, the second mode contains a state $\ket{\psi_0}$ }
\label{fig:figure1}
	\end{figure}

Here we propose a fundamentally different method based on quantum teleportation for realizing quantum enhanced phase measurements. The essence of our proposal is to repeatedly teleport back the probe to coherently apply the phase shift multiple times (see Fig.~\ref{fig:figure1}). This circumvents the need for physically redirecting a probe state to the same phase shift multiple times and allows to keep the entangled resources separate from the potentially lossy phase shifting system. We describe how this protocol can be implemented with current technology of continuous variable teleportation using two-mode squeezed states and an initial coherent state as a probe.   

In the general setup, we consider some initial probe in a state $\ket{\psi_0}$ which is subject to an unknown phase shift described by a unitary $U(\phi)=e^{i\phi\hat{n}}$, where $\hat{n}$ is the number operator. The goal is to estimate the phase $\phi$. After the interaction, an entangled state is used to teleport the output state $U(\phi)\ket{\psi_0}$ back to interact with the phase shift again. This process is then iterated $m$ times. If the teleportation is perfect, this would correspond to the transformation $\ket{\psi_0}\to \left(U(\phi)\right)^{(m+1)}\ket{\psi_0}=U((m+1)\phi)\ket{\psi_0}$ of the input state where $m$ is the number of teleportations. By coherently applying the phase $(m+1)$ times, the signal can have both super resolution and super sensitivity since it will now depend on $(m+1)\phi$ instead of just $\phi$~\cite{giovannetti2006quantum,vanDam2007}.    

As a physical realization of this protocol, we consider the setup illustrated in Fig.~\ref{fig:figure1} where consecutive two-mode squeezed vacuum states are supplied by interferring the output of two single mode squeezed vacuum sources on a balanced beam splitter~\cite{Furusawa1998}. One mode is delayed by some time $T$ and will subsequently be subject to an unknown phase shift described by the unitary $U(\phi)$. Feedback based on previous measurements is applied before the phase shift. The delay $T$ is chosen such that the phase shifted mode can be interfered with the first mode of the next two-mode squeezed vacuum state on a balanced beam splitter before measurement. This setup is inspired by Ref.~\cite{Yokoyama2013} where the continuous generation of continuous variable cluster states is demonstrated. We choose the measurements and the feedback such that the CV teleportation protocol of Ref.~\cite{Braunstein1998}, is realized. In this teleportation protocol, the momentum quadrature of one of the output modes and the position quadrature of the other is measured, which can be achieved with homodyne detection. The feedback then consists of a displacement of the momentum and position quadratures based on the measurement outcomes. 
For perfect teleportation, infinitely many photons are, in principle, needed in the two-mode squeezed vacuum states. The number of photons actually obtaining the phase shift will nonetheless only depend on the initial input state. For situations where the phase shift is obtained by interaction with some fragile system, the effective number of  probe photons actually interacting with the system will be $\sim (m+1) n_{0}$ where $n_{0}$ is the number of photons in the initial state. We will show that Heisenberg limited sensitivity in terms of probe photons can be reached with a simple coherent state as input state and two-mode squeezed states containing on average $\sim m$ photons. Furthermore, the phase resolution can be enhanced by a factor of $m+1$.  

We consider a coherent state $|\alpha\rangle$ with $\avg{\hat{p}}=\alpha$ as the initial probe state $\ket{\psi_0}$. In the setup in Fig.~\ref{fig:figure1}, we can think of displacing the initial vacuum mode of the lower arm before the first measurements. After the interaction of $U(\phi)$, the  state will be $\ket{\alpha e^{i\phi}}$. This state is now teleported back to the second mode of the first two-mode squeezed vacuum state following the CV protocol of Ref.~\cite{Braunstein1998}. The two-mode squeezed vacuum state has squeezing parameter $r$ such that $\avg{\left(\hat{x}_2-\hat{x}_3\right)^2}=e^{-2r}/2$ , where $\hat{x}_2$, $\hat{x}_3$ are the position quadratures for the two modes. The first mode of the two-mode squeezed vacuum state is mixed with the probe state on a balanced beamsplitter. The output modes of the beamsplitter have position quadratures $\hat{x}'_{1}=(\hat{x}_1+\hat{x}_2)/\sqrt{2}$ and $\hat{x}'_{2}=(\hat{x}_1-\hat{x}_2)/\sqrt{2}$ with similar expressions for the momentum quadratures. Here $\hat{x}_1$ is the position quadrature of the probe state. The quadratures $\hat{p}'_1$ and $\hat{x}'_2$ are now measured giving measurement outcomes $\{p'_1,x'_2\}$. Finally, a feedback implements the displacements $\hat{x}_3\to\hat{x}'_3=\hat{x}_3+g_x\sqrt{2}x'_2$ and $\hat{p}_3\to\hat{p}'_3=\hat{p}_3+g_p\sqrt{2}p'_1$, which concludes the teleportation protocol of Ref.~\cite{Braunstein1998}.

\begin{figure} [t] 
\centering
\subfloat {\label{fig:figure2a}\includegraphics[width=0.45\textwidth]{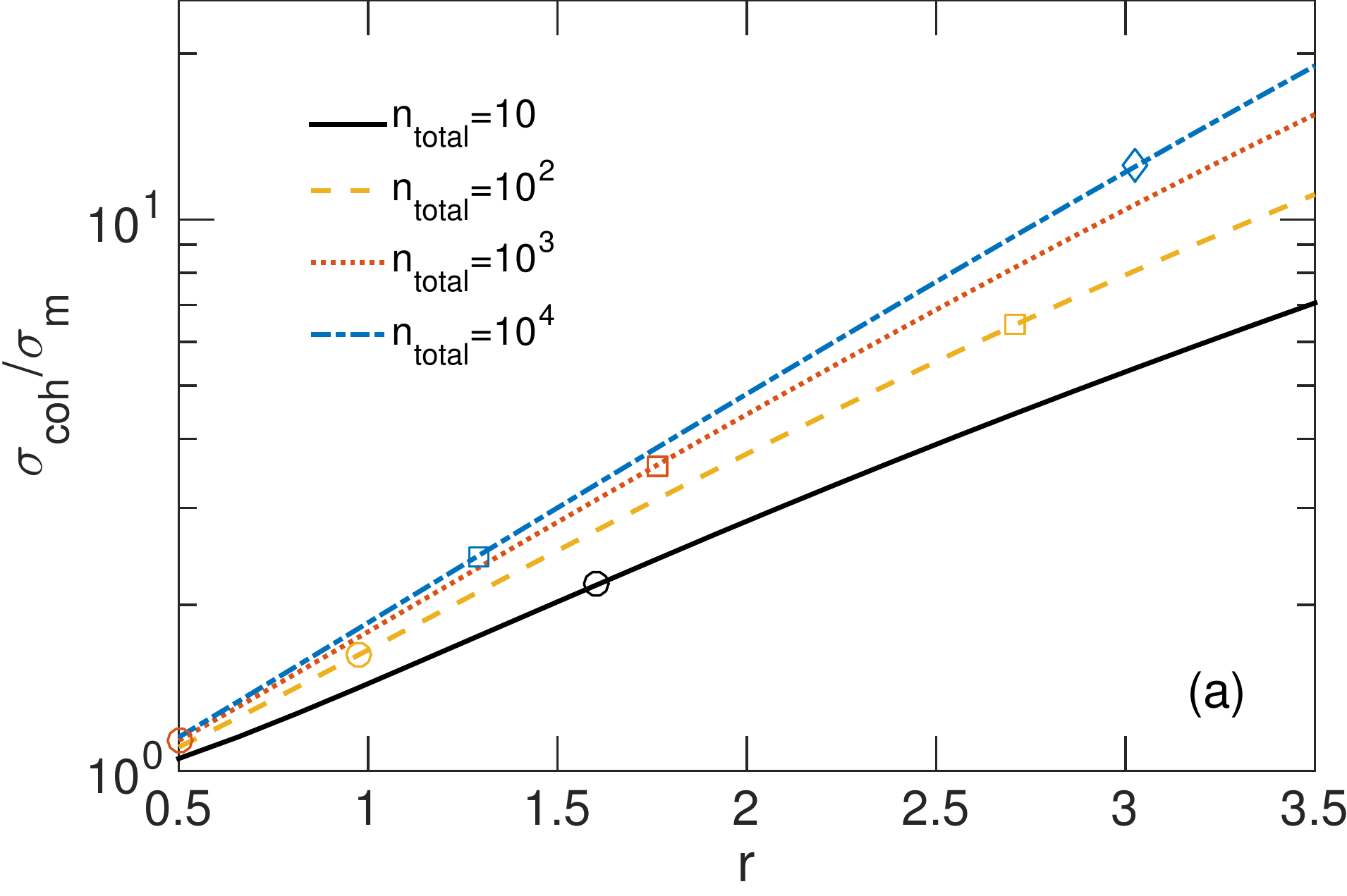}} \\
\subfloat{\label{fig:figure2b}\includegraphics[width=0.45\textwidth]{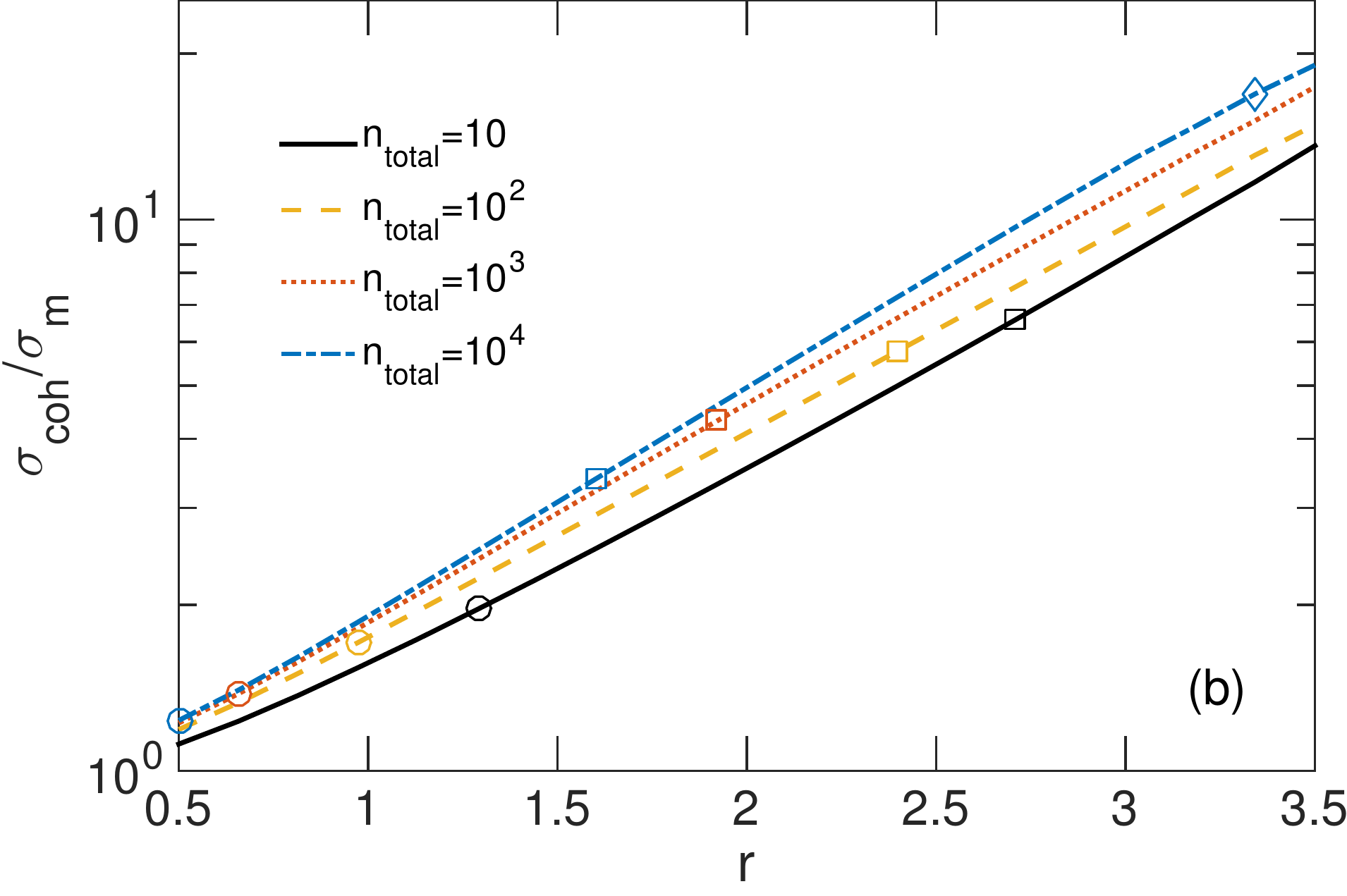}}
\caption{Maximum gain in sensitivity by using the teleportation scheme compared to a classical coherent state protocol for limited amount of squeezing ($r$) and fixed average number of total photons $\bar{n}_{\text{total}}$. We have assumed that $|\cos((m+1)\phi)|\approx1$. The performance is better for high $\bar{n}_{\text{total}}$ because here the photons added to the probe states by imperfect teleportation has smaller weight compared to the extra noise added to the quadrature. We have assumed gains of $g_x=g_p=1$ in (a) while we have numerically optimized the gains in (b). It is seen that in the limit $\bar{n}_{\text{total}}\lesssim e^{2r}$ the optimal gains are different from $g_x=g_p=1$. The optimal number of teleportations $m$ found in the optimizations are indicated with red circles, squares and diamonds. These indicate the transitions to $m\geq10,100$, and $1000$, respectively on the curves. }
\label{fig:figure2}
\end{figure}

The feedback displaces the quadratures such that the teleported state, $\ket{\psi_{1}}$ will be close to $\ket{\alpha e^{i\phi}}$. The quality of the teleportation will depend on the amount of squeezing contained in the two-mode squeezed vacuum state and the feedback strength quantified by the gains $g_x$ and $g_p$. In the limit of high squeezing, perfect teleportation is obtained for $g_x=g_p=1$. The protocol now repeats itself $m$ times corresponding to $m$ teleportations being performed. Finally, the position quadrature ($\hat{x}_m$) of the final state, $\ket{\psi_{\text{m}}}$ is measured. Assuming gains of $g_x=g_p=1$, the mean and variance of $\hat{x}_m$ is
\begin{eqnarray} 
\avg{\hat{x}_m}&=&\bra{\psi_{\text{m}}}\hat{x}_m\ket{\psi_{\text{m}}}=\alpha\sin((m+1)\phi) \label{eq:mean} \\
\text{Var}(\hat{x}_m)&=&\bra{\psi_{\text{m}}}\left(\hat{x}_m^2-\avg{\hat{x}_m}^2\right)\ket{\psi_{\text{m}}}=\frac{1+2me^{-2r}}{4}. \label{eq:var}
\end{eqnarray}
It is clear from Eq.~(\ref{eq:mean}) that the signal exhibits super-resolution in $\phi$ by a factor of $(m+1)$. The sensitivity of the measurement can be quantified as~\cite{Rafal2015}
\begin{equation} \label{eq:sensitivity}
\sigma_{m}=\frac{\sqrt{\text{Var}(\hat{x}_m)}}{|\delta\avg{\hat{x}_m}/\delta\phi|}=\frac{\sqrt{1+2me^{-2r}}}{2(m+1)\alpha|\cos((m+1)\phi)|}. 
\end{equation} 
Note that the sensitivity exhibits a linear decrease in the number of teleportations $m$ as long as $|\cos((m+1)\phi)|\approx1$ and the squeezing is sufficiently strong such that $2me^{-2r}\ll1$. If $m$ consecutive coherent states $\ket{\alpha}$ had been employed, the sensitivity would have a scaling of $\propto1/(\sqrt{m}\alpha)$.  
The average number of probe photons, $\bar{n}_{m}$ contained in the state $\ket{\psi_{m}}$ is  
\begin{equation}
\bar{n}_{m}=\alpha^2+me^{-2r}, 
\end{equation}
thus the total average number of probe photons that have interacted with the phase shift operator will be
\begin{equation} \label{eq:photonnumber}
\bar{n}_{\text{total}}=\sum_{i=0}^{m}\bar{n}_{i}=(m+1)\alpha^2+\frac{1}{2}m(m+1)e^{-2r}. 
\end{equation}
If the coherent state contains one photon ($\alpha=1$) on average, we have that $\bar{n}_{\text{total}}=(m+1)(1+\frac{1}{2}me^{-2r})$ and the sensitivity is
\begin{equation}
\sigma_{m}=\frac{\sqrt{\left(1+\frac{1}{2}me^{-2r}\right)^2\left(1+2me^{-2r}\right)}}{2\bar{n}_{\text{total}}|\cos((m+1)\phi)|}.  
\end{equation}
Thus if $me^{-2r}\ll1$, the sensitivity exhibits Heisenberg scaling in the number of photons for $|\cos((m+1)\phi)|\approx1$. This sensitivity is similar to what could be obtained using NOON states of $(m+1)$ photons and single photon detection and expresses the ultimate scaling allowed by quantum mechanics~\cite{giovannetti2004quantum}.  
	
One of the dominant experimental limitation of the proposed protocol will arguably be the amount of squeezing in the two-mode squeezed vacuum states. This will limit how many teleportations can be performed before the extra noise from the imperfect teleportations will dominate the signal. We therefore consider what the optimum strategy is given a constraint on the amount of squeezing. We consider both a limitation on the amount of squeezing and on the total average number of photons that can interact with the phase shift system. We then optimize over the number of teleportations $m$ and the size of the coherent probe state $\alpha$, to find the strategy that provides the maximum sensitivity for these limitations. Furthermore, we also allow for arbitrary gains $g_x$ and $g_p$. The result of the optimization is shown in Fig.~\ref{fig:figure2} where we illustrate the performance relative to a standard coherent state protocol with matched average photon number. For such an approach, the sensitivity is simply $\sigma_{\text{coh}}=1/(2\sqrt{\bar{n}_{\text{total}}}|\cos(\phi)|)$ where $\bar{n}_{\text{total}}$ is the average number of probe photons. For $|\cos(\phi)|\approx1$, the coherent state approach exhibits sensitivity at the SQL. Fig.~\ref{fig:figure2} shows the two effects of the imperfect teleportation; noise is added in the $\hat{x}$-quadrature (see Eq.~(\ref{eq:var})) and more photons are added to the probe state (see Eq.~(\ref{eq:photonnumber})). In the minimization, the error from the extra photons added by an imperfect teleportation has smaller weight for higher $n_{\text{total}}$. In the limit where $n_{\text{total}}\gg e^{2r}$, the enhancement is $\sim e^{r}/\sqrt{2}$ and equal gains of $g_x=g_p=1$ are optimal. This is the limit where the extra photons added to the probe state does not have any significant effect on the optimum performance. We note that a similar enhancement in sensitivity could be obtained by using a squeezed coherent state as probe~\cite{Caves1981,Bondurant1984}. For such protocols, the squeezed photons, however, interact with the phase shift system, which is not the case here. Consequently, this protocol also works in the limit $n_{\text{total}}\ll e^{2r}$ where an enhancement of $\sim\!\!\left(n_{\text{total}}e^{2r}/2\right)^{\frac{1}{4}}$ can be obtained for $g_x=g_p=1$. Note that our numerical optimization shows that larger enhancement can also be obtained for optimized gains in this limit (see Fig.~\ref{fig:figure2}).    
 
One of the technological challenges of using highly entangled quantum states for enhanced phase measurements is that they are very fragile to losses. Multi-pass protocols share this fragility since losses grow exponentially with the number of passes through the sample~\cite{Demkowicz2010}. This means that if the losses are too high, the sensitivity enhancement of the multi-pass protocol proposed here will vanish. Note, however that while approaches based on NOON states rely on single photon detection this protocol is based on homodyne detection, which in practice is much more efficient. Since imperfect photon detection will add to the overall loss this means that the effective loss may be substantially reduced with this protocol. 
	
\begin{figure} [t] 
\centering
\subfloat {\label{fig:figure3a}\includegraphics[width=0.45\textwidth]{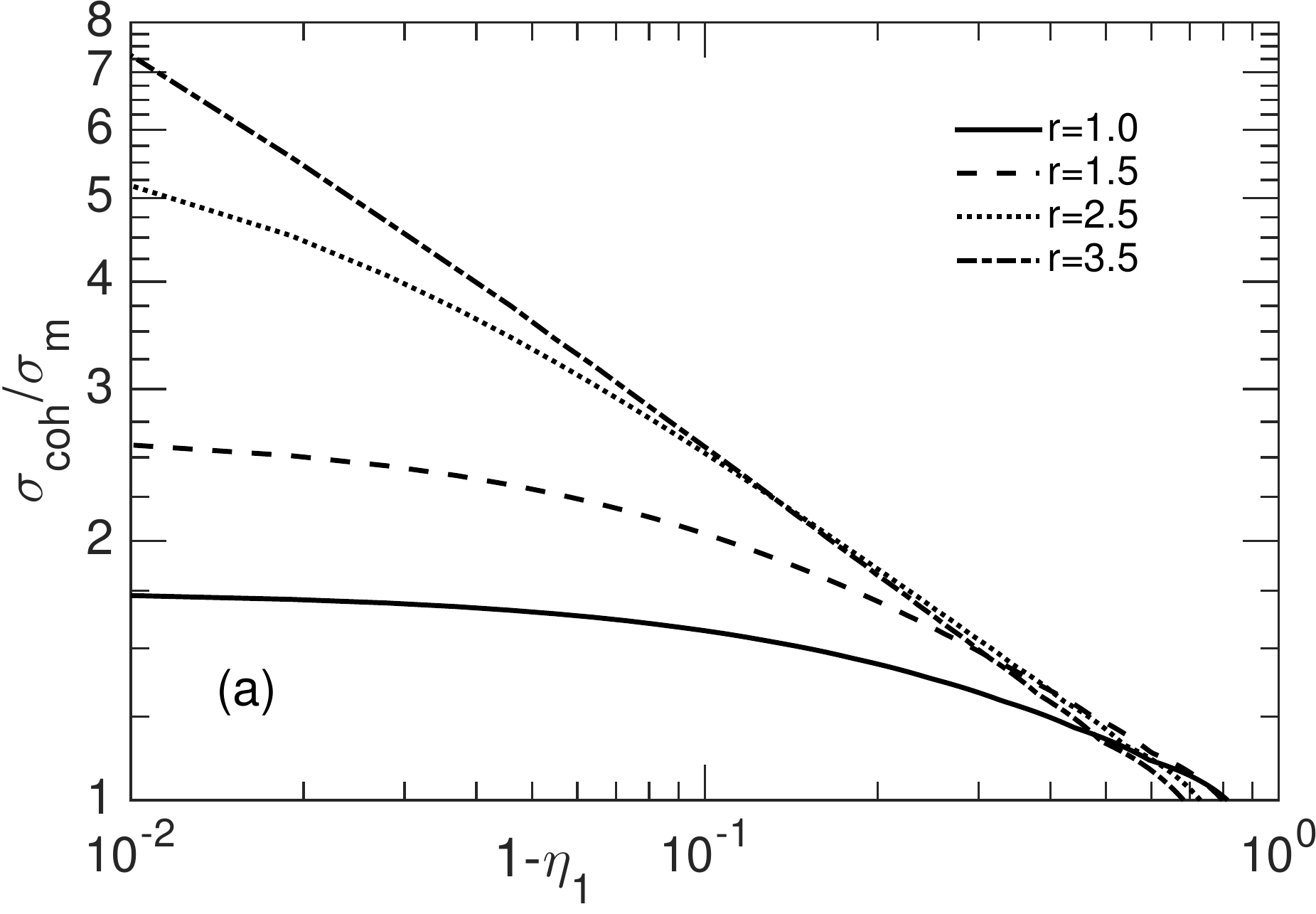}} \\
\subfloat{\label{fig:figure3b}\includegraphics[width=0.45\textwidth]{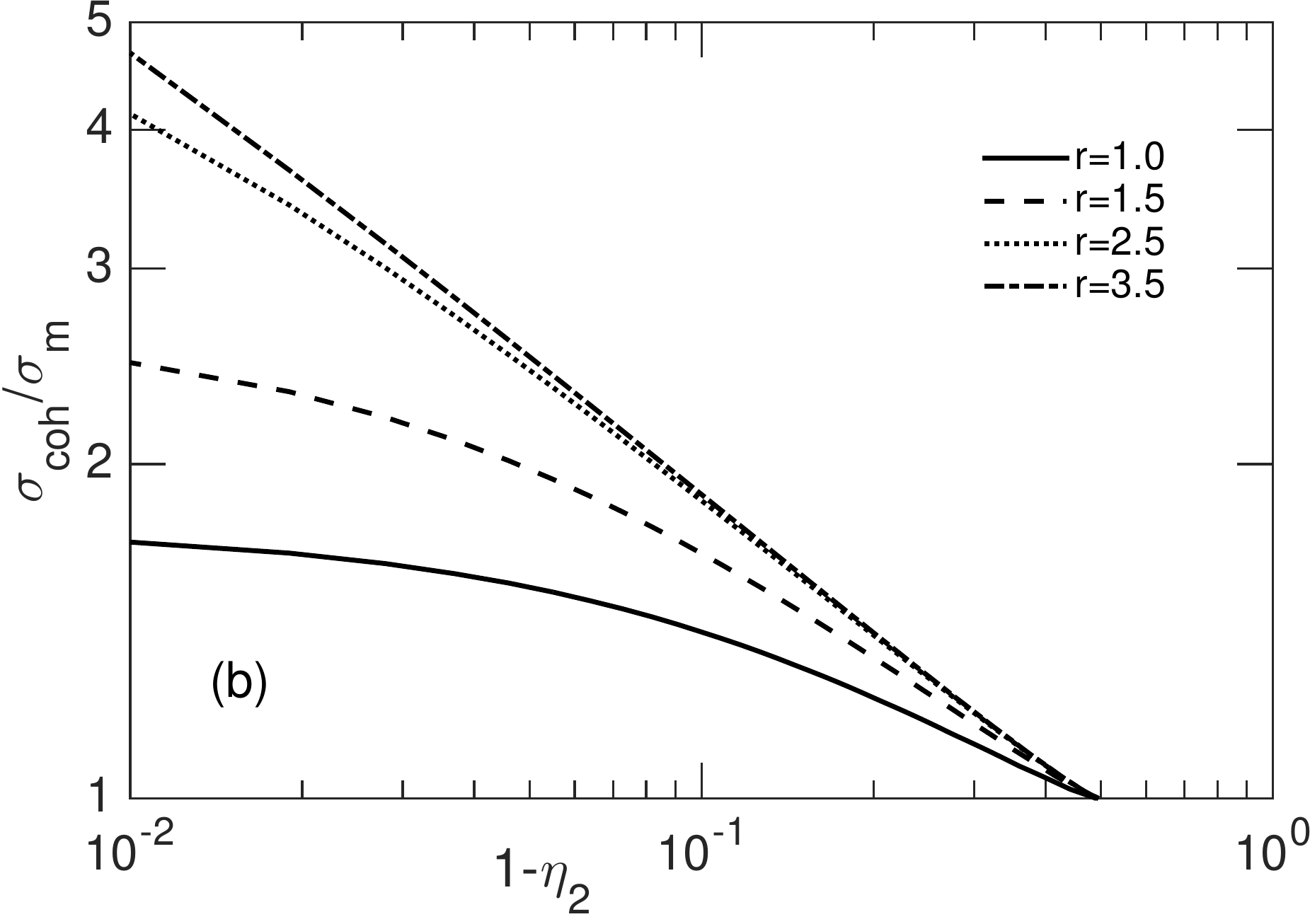}}
\caption{Maximum gain in sensitivity by using the teleportation scheme compared to a classical coherent state protocol for limited amount of squeezing ($r$) and fixed average number of total photons $\bar{n}_{\text{total}}=100$ in the presence of (a) loss on the probe state ($\eta_1<1$, $\eta_2=1$) and (b) loss on the two-mode squeezed vacuum state ($\eta_1=1$, $\eta_2<1$). We have assumed that $|\cos((m+1)\phi)|\approx1$. }
\label{fig:figure3}
\end{figure}

We investigate the performance of the proposed protocol in the presence of both losses acting on the probe state corresponding to a lossy phase shift system and losses acting on the two-mode squeezed vacuum states.  We model the losses with fictitious beamsplitters where the unused output port is traced out. To model the lossy phase shift system a fictitious beamsplitter of transmission $\eta_1$ is inserted after the phase shift $U(\phi)$ (see Fig.~\ref{fig:figure1}). For the loss in the two-mode squeezed vacuum state, fictitious beamsplitters both with transmission $\eta_2$ are inserted for each of the modes. For simplicity, we have assumed equal losses for both modes. Assuming equal gains of $g_x=g_p=1$, the signal and sensitivity after $m$ teleportations for $\{\eta_1,\eta_2\}<1$ is
\begin{eqnarray}
\avg{\hat{x}}_m&=&\alpha\eta_1^{\frac{m+1}{2}}\sin((m+1)\phi) \label{eq:losssignal} \\
\sigma_m&=&\frac{\sqrt{1+2\eta_1\frac{1-\eta_1^m}{1-\eta_1}\left(\eta_2e^{-2r}+1-\eta_2\right)}}{2(m+1)\alpha\eta_1^{\frac{m+1}{2}}|\cos((m+1)\phi)|}.  \label{eq:sensitivityloss}
\end{eqnarray}
As expected, the loss on the probe state ($\eta_1$) enters in the expression for the sensitivity exponentially in $m$, while loss on the two-mode squeezed vacuum states ($\eta_2$) only has a linear effect in $m$. The effect of $\eta_2<1$ on the sensitivity is equivalent to having a limited squeezing of $r_{\text{lim}}=-\frac{1}{2}\ln\left(\eta_2e^{-2r}+1-\eta_2\right)$.  This also holds when considering the average number of total probe photons incident on the phase shift system.  For $m$ teleportations and gains of $g_x=g_p=1$, we have that
\begin{eqnarray}
\bar{n}_{\text{total}}&=&\frac{1-\eta_1^{m+1}}{1-\eta_1}\alpha^2\nonumber \\
&&+\frac{m(1-\eta_1)-\eta_1(1-\eta_1^m)}{(1-\eta_1)^2}\left(\eta_2e^{-2r}+1-\eta_2\right). \qquad \label{eq:photonloss}
\end{eqnarray}
Note that by taking the limit $\eta_1\to1$ for $\eta_2=1$, Eqs.~(\ref{eq:losssignal})-(\ref{eq:photonloss}) reduces to Eqs.~(\ref{eq:mean}),~(\ref{eq:sensitivity}), and (\ref{eq:photonnumber}). If excess noise on the squeezed states is included by mixing in thermal states of average photon number $\bar{n}$ instead of vacuum in the fictitious beam splitters ($\eta_2$), one would have that $r_{\text{lim}}=-\frac{1}{2}\ln\left(\eta_2e^{-2r}+(1+2\bar{n})(1-\eta_2\right))$. We will assume that $\bar{n}\ll1$ such that excess noise can be neglected. To see the effect of finite losses, we again compare the protocol to the simple coherent strategy for which $\sigma_{\text{coh}}=1/(2\sqrt{\eta_1 n_{\text{total}}}|\cos(\phi)|)$ in the presence of loss. The result of the optimization is shown in Fig.~\ref{fig:figure3}. While a small improvement was found by optimizing the gains, near optimal performance is reached for $g_x=g_p=1$. The error from losses in the two-mode squeezed vacuum state limits the gain in the same way as finite squeezing does for the lossless case. Consequently, when these losses dominate the error, the enhancement is $\sim1/\sqrt{2(1-\eta_2)}$ and no enhancement is possible for $\eta_2\sim1/2$. When losses in the probe state limit the enhancement, the optimum performance is effectively found as a tradeoff between the $\sqrt{m+1}$ enhancement due to the teleportation and the exponential reduction due to the loss. As a result, we find that the enhancement is $~\sim\sqrt{2/(3(1-\eta_1))}$ and no enhancement is possible for $\eta_1\sim1/3$. We note that while losses quickly reduce the enhancement, the scheme still exhibits enhanced sensitivity compared to the standard coherent state probe even for substantial losses. 

Our method can be easily extended to a multi-mode scheme to demonstrate Heisenberg-limited imaging. This can be realized by replacing the single-mode teleportation scheme with a multi-mode scheme in which multiple higher-order spatial modes are simultaneously teleported~\cite{Sokolov2001}. Using such a multi-mode approach, sub-shot noise and eventually Heisenberg-limited microscopy can be realized

In conclusion, we have shown how both super-sensitivity and super-resolution can be obtained for an optical phase measurement using continuous variable quantum teleportation based on two-mode squeezed vacuum states. For negligible losses, the protocol can exhibit Heisenberg limited sensitivity ($\sim1/N$) for squeezing $2Ne^{-2r}\ll1$ and increase the resolution by a factor of $N$, where $N$ is the number of probes. While this is equivalent to the enhancement possible with $N$-photon NOON states and single photon detection~\cite{Rafal2015}, the protocol proposed here relies on homodyne detection, which generally is more efficient than single photon detection. As a consequence of the super-resolution, the phase to be estimated should, in principle, be localized within a window of $1/N$ to reach the Heisenberg limit as for a NOON or GHZ state approach~\cite{giovannetti2004quantum,pan2012}. However, methods developed to estimate arbitrary phases~\cite{Berni2015,Xiang2010}, in particular, for NOON~\cite{mitchell2005} and GHZ states~\cite{kessler2014} might also be employed in a straightforward way to this scheme. The latter method uses GHZ states of different sizes in order to estimate the digits of the phase allowing for arbitrary phase estimation~\cite{kessler2014}. The same technique could be employed here by operating with different number of teleportations before readout, which effectively corresponds to sending entangled states of varying sizes. We have also studied the effect of photon loss on the scheme both for loss in the two-mode squeezed vacuum states used for teleportation (limits the effective squeezing) and for loss on the probes corresponding to a lossy phase shift system. While loss quickly reduces the performance, the protocol may still provide super-sensitivity for loss on the order of several percent. For an effective squeezing of 13 dB ($r=1.5$), a 6 dB enhancement of the sensitivity ($\sigma^2$) may be obtained with $100$ probe photons and 10\% loss in the phase shift system. 

While the specific protocol studied here employed CV teleportation of a coherent state with two-mode squeezed vacuum states, the generic setup of teleporting back a probe state to interact with the phase shift system multiple times may be extended to other scenarios. In particular, non-Gaussian states such as photon subtracted two-mode squeezed states~\cite{Kaushik2015} may be considered for enhanced teleportation performance or different probe states providing better single-shot estimation~\cite{Bouchard2017}. A discrete variable variant of the protocol could also be envisioned using 1D-cluster states emitted by single quantum emitters~\cite{Lindner2009}. Here every second qubit could probe the phase shift while the remaining qubits are used to teleport the phase information on from probe to probe. 

\begin{acknowledgments}
We would like to thank Matthias Christandl for valuable feedback on the manuscript and helpful discussions. We also acknowledge funding from Center for Macroscopic Quantum States (bigQ DNRF142) and Qubiz - Quantum Innovation Center. JB acknowledges financial support from the European Research Council (ERC Grant Agreements no 337603), the Danish Council for Independent Research (Sapere Aude), VILLUM FONDEN via the QMATH Centre of Excellence (Grant No. 10059).  
	\end{acknowledgments}

	\clearpage


\begin{thebibliography}{36}%
\makeatletter
\providecommand \@ifxundefined [1]{%
 \@ifx{#1\undefined}
}%
\providecommand \@ifnum [1]{%
 \ifnum #1\expandafter \@firstoftwo
 \else \expandafter \@secondoftwo
 \fi
}%
\providecommand \@ifx [1]{%
 \ifx #1\expandafter \@firstoftwo
 \else \expandafter \@secondoftwo
 \fi
}%
\providecommand \natexlab [1]{#1}%
\providecommand \enquote  [1]{``#1''}%
\providecommand \bibnamefont  [1]{#1}%
\providecommand \bibfnamefont [1]{#1}%
\providecommand \citenamefont [1]{#1}%
\providecommand \href@noop [0]{\@secondoftwo}%
\providecommand \href [0]{\begingroup \@sanitize@url \@href}%
\providecommand \@href[1]{\@@startlink{#1}\@@href}%
\providecommand \@@href[1]{\endgroup#1\@@endlink}%
\providecommand \@sanitize@url [0]{\catcode `\\12\catcode `\$12\catcode
  `\&12\catcode `\#12\catcode `\^12\catcode `\_12\catcode `\%12\relax}%
\providecommand \@@startlink[1]{}%
\providecommand \@@endlink[0]{}%
\providecommand \url  [0]{\begingroup\@sanitize@url \@url }%
\providecommand \@url [1]{\endgroup\@href {#1}{\urlprefix }}%
\providecommand \urlprefix  [0]{URL }%
\providecommand \Eprint [0]{\href }%
\providecommand \doibase [0]{http://dx.doi.org/}%
\providecommand \selectlanguage [0]{\@gobble}%
\providecommand \bibinfo  [0]{\@secondoftwo}%
\providecommand \bibfield  [0]{\@secondoftwo}%
\providecommand \translation [1]{[#1]}%
\providecommand \BibitemOpen [0]{}%
\providecommand \bibitemStop [0]{}%
\providecommand \bibitemNoStop [0]{.\EOS\space}%
\providecommand \EOS [0]{\spacefactor3000\relax}%
\providecommand \BibitemShut  [1]{\csname bibitem#1\endcsname}%
\let\auto@bib@innerbib\@empty
\bibitem [{\citenamefont {Giovannetti}\ \emph {et~al.}(2004)\citenamefont
  {Giovannetti}, \citenamefont {Lloyd},\ and\ \citenamefont
  {Maccone}}]{giovannetti2004quantum}%
  \BibitemOpen
  \bibfield  {author} {\bibinfo {author} {\bibfnamefont {V.}~\bibnamefont
  {Giovannetti}}, \bibinfo {author} {\bibfnamefont {S.}~\bibnamefont {Lloyd}},
  \ and\ \bibinfo {author} {\bibfnamefont {L.}~\bibnamefont {Maccone}},\
  }\href@noop {} {\bibfield  {journal} {\bibinfo  {journal} {Science}\ }\textbf
  {\bibinfo {volume} {306}},\ \bibinfo {pages} {1330} (\bibinfo {year}
  {2004})}\BibitemShut {NoStop}%
\bibitem [{\citenamefont {Giovannetti}\ \emph {et~al.}(2011)\citenamefont
  {Giovannetti}, \citenamefont {Lloyd},\ and\ \citenamefont
  {Maccone}}]{giovannetti2011advances}%
  \BibitemOpen
  \bibfield  {author} {\bibinfo {author} {\bibfnamefont {V.}~\bibnamefont
  {Giovannetti}}, \bibinfo {author} {\bibfnamefont {S.}~\bibnamefont {Lloyd}},
  \ and\ \bibinfo {author} {\bibfnamefont {L.}~\bibnamefont {Maccone}},\
  }\href@noop {} {\bibfield  {journal} {\bibinfo  {journal} {Nature Photonics}\
  }\textbf {\bibinfo {volume} {5}},\ \bibinfo {pages} {222} (\bibinfo {year}
  {2011})}\BibitemShut {NoStop}%
\bibitem [{\citenamefont {Escher}\ \emph {et~al.}(2011)\citenamefont {Escher},
  \citenamefont {de~Matos~Filho},\ and\ \citenamefont
  {Davidovich}}]{Escher2011}%
  \BibitemOpen
  \bibfield  {author} {\bibinfo {author} {\bibfnamefont {B.~M.}\ \bibnamefont
  {Escher}}, \bibinfo {author} {\bibfnamefont {R.~L.}\ \bibnamefont
  {de~Matos~Filho}}, \ and\ \bibinfo {author} {\bibfnamefont {L.}~\bibnamefont
  {Davidovich}},\ }\href {http://dx.doi.org/10.1038/nphys1958} {\bibfield
  {journal} {\bibinfo  {journal} {Nature Physics}\ }\textbf {\bibinfo {volume}
  {7}},\ \bibinfo {pages} {406 EP } (\bibinfo {year} {2011})}\BibitemShut
  {NoStop}%
\bibitem [{\citenamefont {Degen}\ \emph {et~al.}(2017)\citenamefont {Degen},
  \citenamefont {Reinhard},\ and\ \citenamefont
  {Cappellaro}}]{Deganreview2017}%
  \BibitemOpen
  \bibfield  {author} {\bibinfo {author} {\bibfnamefont {C.~L.}\ \bibnamefont
  {Degen}}, \bibinfo {author} {\bibfnamefont {F.}~\bibnamefont {Reinhard}}, \
  and\ \bibinfo {author} {\bibfnamefont {P.}~\bibnamefont {Cappellaro}},\
  }\href {\doibase 10.1103/RevModPhys.89.035002} {\bibfield  {journal}
  {\bibinfo  {journal} {Rev. Mod. Phys.}\ }\textbf {\bibinfo {volume} {89}},\
  \bibinfo {pages} {035002} (\bibinfo {year} {2017})}\BibitemShut {NoStop}%
\bibitem [{\citenamefont {Pan}\ \emph {et~al.}(2012)\citenamefont {Pan},
  \citenamefont {Chen}, \citenamefont {Lu}, \citenamefont {Weinfurter},
  \citenamefont {Zeilinger},\ and\ \citenamefont {\ifmmode~\dot{Z}\else
  \.{Z}\fi{}ukowski}}]{pan2012}%
  \BibitemOpen
  \bibfield  {author} {\bibinfo {author} {\bibfnamefont {J.-W.}\ \bibnamefont
  {Pan}}, \bibinfo {author} {\bibfnamefont {Z.-B.}\ \bibnamefont {Chen}},
  \bibinfo {author} {\bibfnamefont {C.-Y.}\ \bibnamefont {Lu}}, \bibinfo
  {author} {\bibfnamefont {H.}~\bibnamefont {Weinfurter}}, \bibinfo {author}
  {\bibfnamefont {A.}~\bibnamefont {Zeilinger}}, \ and\ \bibinfo {author}
  {\bibfnamefont {M.}~\bibnamefont {\ifmmode~\dot{Z}\else \.{Z}\fi{}ukowski}},\
  }\href {\doibase 10.1103/RevModPhys.84.777} {\bibfield  {journal} {\bibinfo
  {journal} {Rev. Mod. Phys.}\ }\textbf {\bibinfo {volume} {84}},\ \bibinfo
  {pages} {777} (\bibinfo {year} {2012})}\BibitemShut {NoStop}%
\bibitem [{\citenamefont {Demkowicz-Dobrza{\'n}ski}\ \emph
  {et~al.}(2015)\citenamefont {Demkowicz-Dobrza{\'n}ski}, \citenamefont
  {Jarzyna},\ and\ \citenamefont {Ko{\l}ody{\'n}ski}}]{Rafal2015}%
  \BibitemOpen
  \bibfield  {author} {\bibinfo {author} {\bibfnamefont {R.}~\bibnamefont
  {Demkowicz-Dobrza{\'n}ski}}, \bibinfo {author} {\bibfnamefont
  {M.}~\bibnamefont {Jarzyna}}, \ and\ \bibinfo {author} {\bibfnamefont
  {J.}~\bibnamefont {Ko{\l}ody{\'n}ski}},\ }\href {\doibase
  https://doi.org/10.1016/bs.po.2015.02.003} {\bibfield  {journal} {\bibinfo
  {journal} {Progress in Optics}\ }\bibinfo {series} {Progress in Optics},\
  \textbf {\bibinfo {volume} {60}},\ \bibinfo {pages} {345 } (\bibinfo {year}
  {2015})}\BibitemShut {NoStop}%
\bibitem [{\citenamefont {Bouchard}\ \emph {et~al.}(2017)\citenamefont
  {Bouchard}, \citenamefont {de~la Hoz}, \citenamefont {Bj\"{o}rk},
  \citenamefont {Boyd}, \citenamefont {Grassl}, \citenamefont {Hradil},
  \citenamefont {Karimi}, \citenamefont {Klimov}, \citenamefont {Leuchs},
  \citenamefont {\v{R}eh\'{a}\v{c}ek},\ and\ \citenamefont
  {S\'{a}nchez-Soto}}]{Bouchard2017}%
  \BibitemOpen
  \bibfield  {author} {\bibinfo {author} {\bibfnamefont {F.}~\bibnamefont
  {Bouchard}}, \bibinfo {author} {\bibfnamefont {P.}~\bibnamefont {de~la Hoz}},
  \bibinfo {author} {\bibfnamefont {G.}~\bibnamefont {Bj\"{o}rk}}, \bibinfo
  {author} {\bibfnamefont {R.~W.}\ \bibnamefont {Boyd}}, \bibinfo {author}
  {\bibfnamefont {M.}~\bibnamefont {Grassl}}, \bibinfo {author} {\bibfnamefont
  {Z.}~\bibnamefont {Hradil}}, \bibinfo {author} {\bibfnamefont
  {E.}~\bibnamefont {Karimi}}, \bibinfo {author} {\bibfnamefont {A.~B.}\
  \bibnamefont {Klimov}}, \bibinfo {author} {\bibfnamefont {G.}~\bibnamefont
  {Leuchs}}, \bibinfo {author} {\bibfnamefont {J.}~\bibnamefont
  {\v{R}eh\'{a}\v{c}ek}}, \ and\ \bibinfo {author} {\bibfnamefont {L.~L.}\
  \bibnamefont {S\'{a}nchez-Soto}},\ }\href {\doibase 10.1364/OPTICA.4.001429}
  {\bibfield  {journal} {\bibinfo  {journal} {Optica}\ }\textbf {\bibinfo
  {volume} {4}},\ \bibinfo {pages} {1429} (\bibinfo {year} {2017})}\BibitemShut
  {NoStop}%
\bibitem [{\citenamefont {Frigault}\ \emph {et~al.}(2009)\citenamefont
  {Frigault}, \citenamefont {Lacoste}, \citenamefont {Swift},\ and\
  \citenamefont {Brown}}]{Frigault2009}%
  \BibitemOpen
  \bibfield  {author} {\bibinfo {author} {\bibfnamefont {M.~M.}\ \bibnamefont
  {Frigault}}, \bibinfo {author} {\bibfnamefont {J.}~\bibnamefont {Lacoste}},
  \bibinfo {author} {\bibfnamefont {J.~L.}\ \bibnamefont {Swift}}, \ and\
  \bibinfo {author} {\bibfnamefont {C.~M.}\ \bibnamefont {Brown}},\ }\href
  {\doibase 10.1242/jcs.033837} {\bibfield  {journal} {\bibinfo  {journal}
  {Journal of Cell Science}\ }\textbf {\bibinfo {volume} {122}},\ \bibinfo
  {pages} {753} (\bibinfo {year} {2009})},\ \Eprint
  {http://arxiv.org/abs/http://jcs.biologists.org/content/122/6/753.full.pdf}
  {http://jcs.biologists.org/content/122/6/753.full.pdf} \BibitemShut {NoStop}%
\bibitem [{\citenamefont {Celebrano}\ \emph {et~al.}(2011)\citenamefont
  {Celebrano}, \citenamefont {Kukura}, \citenamefont {Renn},\ and\
  \citenamefont {Sandoghdar}}]{Celebrano2011}%
  \BibitemOpen
  \bibfield  {author} {\bibinfo {author} {\bibfnamefont {M.}~\bibnamefont
  {Celebrano}}, \bibinfo {author} {\bibfnamefont {P.}~\bibnamefont {Kukura}},
  \bibinfo {author} {\bibfnamefont {A.}~\bibnamefont {Renn}}, \ and\ \bibinfo
  {author} {\bibfnamefont {V.}~\bibnamefont {Sandoghdar}},\ }\href
  {http://dx.doi.org/10.1038/nphoton.2010.290} {\bibfield  {journal} {\bibinfo
  {journal} {Nature Photonics}\ }\textbf {\bibinfo {volume} {5}},\ \bibinfo
  {pages} {95 EP } (\bibinfo {year} {2011})}\BibitemShut {NoStop}%
\bibitem [{\citenamefont {Piliarik}\ and\ \citenamefont
  {Sandoghdar}(2014)}]{Piliarik2014}%
  \BibitemOpen
  \bibfield  {author} {\bibinfo {author} {\bibfnamefont {M.}~\bibnamefont
  {Piliarik}}\ and\ \bibinfo {author} {\bibfnamefont {V.}~\bibnamefont
  {Sandoghdar}},\ }\href {http://dx.doi.org/10.1038/ncomms5495} {\bibfield
  {journal} {\bibinfo  {journal} {Nature Communications}\ }\textbf {\bibinfo
  {volume} {5}},\ \bibinfo {pages} {4495 EP } (\bibinfo {year}
  {2014})}\BibitemShut {NoStop}%
\bibitem [{\citenamefont {Mitchell}\ \emph {et~al.}(2004)\citenamefont
  {Mitchell}, \citenamefont {Lundeen},\ and\ \citenamefont
  {Steinberg}}]{Mitchell2004}%
  \BibitemOpen
  \bibfield  {author} {\bibinfo {author} {\bibfnamefont {M.~W.}\ \bibnamefont
  {Mitchell}}, \bibinfo {author} {\bibfnamefont {J.~S.}\ \bibnamefont
  {Lundeen}}, \ and\ \bibinfo {author} {\bibfnamefont {A.~M.}\ \bibnamefont
  {Steinberg}},\ }\href {http://dx.doi.org/10.1038/nature02493} {\bibfield
  {journal} {\bibinfo  {journal} {Nature}\ }\textbf {\bibinfo {volume} {429}},\
  \bibinfo {pages} {161 EP } (\bibinfo {year} {2004})}\BibitemShut {NoStop}%
\bibitem [{\citenamefont {Afek}\ \emph {et~al.}(2010)\citenamefont {Afek},
  \citenamefont {Ambar},\ and\ \citenamefont {Silberberg}}]{Afek2010}%
  \BibitemOpen
  \bibfield  {author} {\bibinfo {author} {\bibfnamefont {I.}~\bibnamefont
  {Afek}}, \bibinfo {author} {\bibfnamefont {O.}~\bibnamefont {Ambar}}, \ and\
  \bibinfo {author} {\bibfnamefont {Y.}~\bibnamefont {Silberberg}},\ }\href
  {\doibase 10.1126/science.1188172} {\bibfield  {journal} {\bibinfo  {journal}
  {Science}\ }\textbf {\bibinfo {volume} {328}},\ \bibinfo {pages} {879}
  (\bibinfo {year} {2010})}\BibitemShut {NoStop}%
\bibitem [{\citenamefont {Wolfgramm}\ \emph {et~al.}(2012)\citenamefont
  {Wolfgramm}, \citenamefont {Vitelli}, \citenamefont {Beduini}, \citenamefont
  {Godbout},\ and\ \citenamefont {Mitchell}}]{Wolfgramm2012}%
  \BibitemOpen
  \bibfield  {author} {\bibinfo {author} {\bibfnamefont {F.}~\bibnamefont
  {Wolfgramm}}, \bibinfo {author} {\bibfnamefont {C.}~\bibnamefont {Vitelli}},
  \bibinfo {author} {\bibfnamefont {F.~A.}\ \bibnamefont {Beduini}}, \bibinfo
  {author} {\bibfnamefont {N.}~\bibnamefont {Godbout}}, \ and\ \bibinfo
  {author} {\bibfnamefont {M.~W.}\ \bibnamefont {Mitchell}},\ }\href
  {http://dx.doi.org/10.1038/nphoton.2012.300} {\bibfield  {journal} {\bibinfo
  {journal} {Nature Photonics}\ }\textbf {\bibinfo {volume} {7}},\ \bibinfo
  {pages} {28 EP } (\bibinfo {year} {2012})}\BibitemShut {NoStop}%
\bibitem [{\citenamefont {Facon}\ \emph {et~al.}(2016)\citenamefont {Facon},
  \citenamefont {Dietsche}, \citenamefont {Grosso}, \citenamefont {Haroche},
  \citenamefont {Raimond}, \citenamefont {Brune},\ and\ \citenamefont
  {Gleyzes}}]{Facon2016}%
  \BibitemOpen
  \bibfield  {author} {\bibinfo {author} {\bibfnamefont {A.}~\bibnamefont
  {Facon}}, \bibinfo {author} {\bibfnamefont {E.-K.}\ \bibnamefont {Dietsche}},
  \bibinfo {author} {\bibfnamefont {D.}~\bibnamefont {Grosso}}, \bibinfo
  {author} {\bibfnamefont {S.}~\bibnamefont {Haroche}}, \bibinfo {author}
  {\bibfnamefont {J.-M.}\ \bibnamefont {Raimond}}, \bibinfo {author}
  {\bibfnamefont {M.}~\bibnamefont {Brune}}, \ and\ \bibinfo {author}
  {\bibfnamefont {S.}~\bibnamefont {Gleyzes}},\ }\href
  {http://dx.doi.org/10.1038/nature18327} {\bibfield  {journal} {\bibinfo
  {journal} {Nature}\ }\textbf {\bibinfo {volume} {535}},\ \bibinfo {pages}
  {262 EP } (\bibinfo {year} {2016})}\BibitemShut {NoStop}%
\bibitem [{\citenamefont {Caves}(1981)}]{Caves1981}%
  \BibitemOpen
  \bibfield  {author} {\bibinfo {author} {\bibfnamefont {C.~M.}\ \bibnamefont
  {Caves}},\ }\href {\doibase 10.1103/PhysRevD.23.1693} {\bibfield  {journal}
  {\bibinfo  {journal} {Phys. Rev. D}\ }\textbf {\bibinfo {volume} {23}},\
  \bibinfo {pages} {1693} (\bibinfo {year} {1981})}\BibitemShut {NoStop}%
\bibitem [{\citenamefont {Collaboration}(2011)}]{Ligo2011}%
  \BibitemOpen
  \bibfield  {author} {\bibinfo {author} {\bibfnamefont {T.~L.~S.}\
  \bibnamefont {Collaboration}},\ }\href {http://dx.doi.org/10.1038/nphys2083}
  {\bibfield  {journal} {\bibinfo  {journal} {Nature Physics}\ }\textbf
  {\bibinfo {volume} {7}},\ \bibinfo {pages} {962 EP } (\bibinfo {year}
  {2011})}\BibitemShut {NoStop}%
\bibitem [{\citenamefont {Aasi}\ and\ \citenamefont {\emph{et
  al.}}(2013)}]{Ligo2013}%
  \BibitemOpen
  \bibfield  {author} {\bibinfo {author} {\bibfnamefont {J.}~\bibnamefont
  {Aasi}}\ and\ \bibinfo {author} {\bibnamefont {\emph{et al.}}},\ }\href
  {http://dx.doi.org/10.1038/nphoton.2013.177} {\bibfield  {journal} {\bibinfo
  {journal} {Nature Photonics}\ }\textbf {\bibinfo {volume} {7}},\ \bibinfo
  {pages} {613 EP } (\bibinfo {year} {2013})}\BibitemShut {NoStop}%
\bibitem [{\citenamefont {Taylor}\ \emph {et~al.}(2013)\citenamefont {Taylor},
  \citenamefont {Janousek}, \citenamefont {Daria}, \citenamefont {Knittel},
  \citenamefont {Hage}, \citenamefont {Bachor},\ and\ \citenamefont
  {Bowen}}]{Taylor2013}%
  \BibitemOpen
  \bibfield  {author} {\bibinfo {author} {\bibfnamefont {M.~A.}\ \bibnamefont
  {Taylor}}, \bibinfo {author} {\bibfnamefont {J.}~\bibnamefont {Janousek}},
  \bibinfo {author} {\bibfnamefont {V.}~\bibnamefont {Daria}}, \bibinfo
  {author} {\bibfnamefont {J.}~\bibnamefont {Knittel}}, \bibinfo {author}
  {\bibfnamefont {B.}~\bibnamefont {Hage}}, \bibinfo {author} {\bibfnamefont
  {H.-A.}\ \bibnamefont {Bachor}}, \ and\ \bibinfo {author} {\bibfnamefont
  {W.~P.}\ \bibnamefont {Bowen}},\ }\href
  {http://dx.doi.org/10.1038/nphoton.2012.346} {\bibfield  {journal} {\bibinfo
  {journal} {Nature Photonics}\ }\textbf {\bibinfo {volume} {7}},\ \bibinfo
  {pages} {229 EP } (\bibinfo {year} {2013})}\BibitemShut {NoStop}%
\bibitem [{\citenamefont {Berni}\ \emph {et~al.}(2015)\citenamefont {Berni},
  \citenamefont {Gehring}, \citenamefont {Nielsen}, \citenamefont
  {H{\"a}ndchen}, \citenamefont {Paris},\ and\ \citenamefont
  {Andersen}}]{Berni2015}%
  \BibitemOpen
  \bibfield  {author} {\bibinfo {author} {\bibfnamefont {A.~A.}\ \bibnamefont
  {Berni}}, \bibinfo {author} {\bibfnamefont {T.}~\bibnamefont {Gehring}},
  \bibinfo {author} {\bibfnamefont {B.~M.}\ \bibnamefont {Nielsen}}, \bibinfo
  {author} {\bibfnamefont {V.}~\bibnamefont {H{\"a}ndchen}}, \bibinfo {author}
  {\bibfnamefont {M.~G.~A.}\ \bibnamefont {Paris}}, \ and\ \bibinfo {author}
  {\bibfnamefont {U.~L.}\ \bibnamefont {Andersen}},\ }\href
  {http://dx.doi.org/10.1038/nphoton.2015.139} {\bibfield  {journal} {\bibinfo
  {journal} {Nature Photonics}\ }\textbf {\bibinfo {volume} {9}},\ \bibinfo
  {pages} {577 EP } (\bibinfo {year} {2015})}\BibitemShut {NoStop}%
\bibitem [{\citenamefont {Sch{\"a}fermeier}\ \emph {et~al.}(2016)\citenamefont
  {Sch{\"a}fermeier}, \citenamefont {Kerdoncuff}, \citenamefont {Hoff},
  \citenamefont {Fu}, \citenamefont {Huck}, \citenamefont {Bilek},
  \citenamefont {Harris}, \citenamefont {Bowen}, \citenamefont {Gehring},\ and\
  \citenamefont {Andersen}}]{Schafermeier2016}%
  \BibitemOpen
  \bibfield  {author} {\bibinfo {author} {\bibfnamefont {C.}~\bibnamefont
  {Sch{\"a}fermeier}}, \bibinfo {author} {\bibfnamefont {H.}~\bibnamefont
  {Kerdoncuff}}, \bibinfo {author} {\bibfnamefont {U.~B.}\ \bibnamefont
  {Hoff}}, \bibinfo {author} {\bibfnamefont {H.}~\bibnamefont {Fu}}, \bibinfo
  {author} {\bibfnamefont {A.}~\bibnamefont {Huck}}, \bibinfo {author}
  {\bibfnamefont {J.}~\bibnamefont {Bilek}}, \bibinfo {author} {\bibfnamefont
  {G.~I.}\ \bibnamefont {Harris}}, \bibinfo {author} {\bibfnamefont {W.~P.}\
  \bibnamefont {Bowen}}, \bibinfo {author} {\bibfnamefont {T.}~\bibnamefont
  {Gehring}}, \ and\ \bibinfo {author} {\bibfnamefont {U.~L.}\ \bibnamefont
  {Andersen}},\ }\href {http://dx.doi.org/10.1038/ncomms13628} {\bibfield
  {journal} {\bibinfo  {journal} {Nature Communications}\ }\textbf {\bibinfo
  {volume} {7}},\ \bibinfo {pages} {13628 EP } (\bibinfo {year}
  {2016})}\BibitemShut {NoStop}%
\bibitem [{not()}]{note1}%
  \BibitemOpen
  \href@noop {} {}\bibinfo {note} {For discussions of super sensitivity versus
  super resolution see Ref.~\cite{giovannetti2011advances} and K. J. Resch
  \emph{et al}, Phys. Rev. Lett. 98, 223601 (2007)}\BibitemShut {NoStop}%
\bibitem [{\citenamefont {Giovannetti}\ \emph {et~al.}(2006)\citenamefont
  {Giovannetti}, \citenamefont {Lloyd},\ and\ \citenamefont
  {Maccone}}]{giovannetti2006quantum}%
  \BibitemOpen
  \bibfield  {author} {\bibinfo {author} {\bibfnamefont {V.}~\bibnamefont
  {Giovannetti}}, \bibinfo {author} {\bibfnamefont {S.}~\bibnamefont {Lloyd}},
  \ and\ \bibinfo {author} {\bibfnamefont {L.}~\bibnamefont {Maccone}},\
  }\href@noop {} {\bibfield  {journal} {\bibinfo  {journal} {Physical Review
  Letters}\ }\textbf {\bibinfo {volume} {96}},\ \bibinfo {pages} {010401}
  (\bibinfo {year} {2006})}\BibitemShut {NoStop}%
\bibitem [{\citenamefont {van Dam}\ \emph {et~al.}(2007)\citenamefont {van
  Dam}, \citenamefont {D'Ariano}, \citenamefont {Ekert}, \citenamefont
  {Macchiavello},\ and\ \citenamefont {Mosca}}]{vanDam2007}%
  \BibitemOpen
  \bibfield  {author} {\bibinfo {author} {\bibfnamefont {W.}~\bibnamefont {van
  Dam}}, \bibinfo {author} {\bibfnamefont {G.~M.}\ \bibnamefont {D'Ariano}},
  \bibinfo {author} {\bibfnamefont {A.}~\bibnamefont {Ekert}}, \bibinfo
  {author} {\bibfnamefont {C.}~\bibnamefont {Macchiavello}}, \ and\ \bibinfo
  {author} {\bibfnamefont {M.}~\bibnamefont {Mosca}},\ }\href {\doibase
  10.1103/PhysRevLett.98.090501} {\bibfield  {journal} {\bibinfo  {journal}
  {Phys. Rev. Lett.}\ }\textbf {\bibinfo {volume} {98}},\ \bibinfo {pages}
  {090501} (\bibinfo {year} {2007})}\BibitemShut {NoStop}%
\bibitem [{\citenamefont {Demkowicz-Dobrza{\'{n}}ski}(2010)}]{Demkowicz2010}%
  \BibitemOpen
  \bibfield  {author} {\bibinfo {author} {\bibfnamefont {R.}~\bibnamefont
  {Demkowicz-Dobrza{\'{n}}ski}},\ }\href {\doibase 10.1134/S1054660X10090306}
  {\bibfield  {journal} {\bibinfo  {journal} {Laser Physics}\ }\textbf
  {\bibinfo {volume} {20}},\ \bibinfo {pages} {1197} (\bibinfo {year}
  {2010})}\BibitemShut {NoStop}%
\bibitem [{\citenamefont {Higgins}\ \emph {et~al.}(2007)\citenamefont
  {Higgins}, \citenamefont {Berry}, \citenamefont {Bartlett}, \citenamefont
  {Wiseman},\ and\ \citenamefont {Pryde}}]{Higgins2007}%
  \BibitemOpen
  \bibfield  {author} {\bibinfo {author} {\bibfnamefont {B.~L.}\ \bibnamefont
  {Higgins}}, \bibinfo {author} {\bibfnamefont {D.~W.}\ \bibnamefont {Berry}},
  \bibinfo {author} {\bibfnamefont {S.~D.}\ \bibnamefont {Bartlett}}, \bibinfo
  {author} {\bibfnamefont {H.~M.}\ \bibnamefont {Wiseman}}, \ and\ \bibinfo
  {author} {\bibfnamefont {G.~J.}\ \bibnamefont {Pryde}},\ }\href
  {http://dx.doi.org/10.1038/nature06257} {\bibfield  {journal} {\bibinfo
  {journal} {Nature}\ }\textbf {\bibinfo {volume} {450}},\ \bibinfo {pages}
  {393 EP } (\bibinfo {year} {2007})}\BibitemShut {NoStop}%
\bibitem [{\citenamefont {Juffmann}\ \emph {et~al.}(2016)\citenamefont
  {Juffmann}, \citenamefont {Klopfer}, \citenamefont {Frankort}, \citenamefont
  {Haslinger},\ and\ \citenamefont {Kasevich}}]{Juffmann2016}%
  \BibitemOpen
  \bibfield  {author} {\bibinfo {author} {\bibfnamefont {T.}~\bibnamefont
  {Juffmann}}, \bibinfo {author} {\bibfnamefont {B.~B.}\ \bibnamefont
  {Klopfer}}, \bibinfo {author} {\bibfnamefont {T.~L.~I.}\ \bibnamefont
  {Frankort}}, \bibinfo {author} {\bibfnamefont {P.}~\bibnamefont {Haslinger}},
  \ and\ \bibinfo {author} {\bibfnamefont {M.~A.}\ \bibnamefont {Kasevich}},\
  }\href {http://dx.doi.org/10.1038/ncomms12858} {\bibfield  {journal}
  {\bibinfo  {journal} {Nature Communications}\ }\textbf {\bibinfo {volume}
  {7}},\ \bibinfo {pages} {12858 EP } (\bibinfo {year} {2016})}\BibitemShut
  {NoStop}%
\bibitem [{\citenamefont {Furusawa}\ \emph {et~al.}(1998)\citenamefont
  {Furusawa}, \citenamefont {S{\o}rensen}, \citenamefont {Braunstein},
  \citenamefont {Fuchs}, \citenamefont {Kimble},\ and\ \citenamefont
  {Polzik}}]{Furusawa1998}%
  \BibitemOpen
  \bibfield  {author} {\bibinfo {author} {\bibfnamefont {A.}~\bibnamefont
  {Furusawa}}, \bibinfo {author} {\bibfnamefont {J.~L.}\ \bibnamefont
  {S{\o}rensen}}, \bibinfo {author} {\bibfnamefont {S.~L.}\ \bibnamefont
  {Braunstein}}, \bibinfo {author} {\bibfnamefont {C.~A.}\ \bibnamefont
  {Fuchs}}, \bibinfo {author} {\bibfnamefont {H.~J.}\ \bibnamefont {Kimble}}, \
  and\ \bibinfo {author} {\bibfnamefont {E.~S.}\ \bibnamefont {Polzik}},\
  }\href {\doibase 10.1126/science.282.5389.706} {\bibfield  {journal}
  {\bibinfo  {journal} {Science}\ }\textbf {\bibinfo {volume} {282}},\ \bibinfo
  {pages} {706} (\bibinfo {year} {1998})},\ \Eprint
  {http://arxiv.org/abs/http://science.sciencemag.org/content/282/5389/706.full.pdf}
  {http://science.sciencemag.org/content/282/5389/706.full.pdf} \BibitemShut
  {NoStop}%
\bibitem [{\citenamefont {Yokoyama}\ \emph {et~al.}(2013)\citenamefont
  {Yokoyama}, \citenamefont {Ukai}, \citenamefont {Armstrong}, \citenamefont
  {Sornphiphatphong}, \citenamefont {Kaji}, \citenamefont {Suzuki},
  \citenamefont {Yoshikawa}, \citenamefont {Yonezawa}, \citenamefont
  {Menicucci},\ and\ \citenamefont {Furusawa}}]{Yokoyama2013}%
  \BibitemOpen
  \bibfield  {author} {\bibinfo {author} {\bibfnamefont {S.}~\bibnamefont
  {Yokoyama}}, \bibinfo {author} {\bibfnamefont {R.}~\bibnamefont {Ukai}},
  \bibinfo {author} {\bibfnamefont {S.~C.}\ \bibnamefont {Armstrong}}, \bibinfo
  {author} {\bibfnamefont {C.}~\bibnamefont {Sornphiphatphong}}, \bibinfo
  {author} {\bibfnamefont {T.}~\bibnamefont {Kaji}}, \bibinfo {author}
  {\bibfnamefont {S.}~\bibnamefont {Suzuki}}, \bibinfo {author} {\bibfnamefont
  {J.-i.}\ \bibnamefont {Yoshikawa}}, \bibinfo {author} {\bibfnamefont
  {H.}~\bibnamefont {Yonezawa}}, \bibinfo {author} {\bibfnamefont {N.~C.}\
  \bibnamefont {Menicucci}}, \ and\ \bibinfo {author} {\bibfnamefont
  {A.}~\bibnamefont {Furusawa}},\ }\href
  {http://dx.doi.org/10.1038/nphoton.2013.287} {\bibfield  {journal} {\bibinfo
  {journal} {Nature Photonics}\ }\textbf {\bibinfo {volume} {7}},\ \bibinfo
  {pages} {982 EP } (\bibinfo {year} {2013})}\BibitemShut {NoStop}%
\bibitem [{\citenamefont {Braunstein}\ and\ \citenamefont
  {Kimble}(1998)}]{Braunstein1998}%
  \BibitemOpen
  \bibfield  {author} {\bibinfo {author} {\bibfnamefont {S.~L.}\ \bibnamefont
  {Braunstein}}\ and\ \bibinfo {author} {\bibfnamefont {H.~J.}\ \bibnamefont
  {Kimble}},\ }\href {\doibase 10.1103/PhysRevLett.80.869} {\bibfield
  {journal} {\bibinfo  {journal} {Phys. Rev. Lett.}\ }\textbf {\bibinfo
  {volume} {80}},\ \bibinfo {pages} {869} (\bibinfo {year} {1998})}\BibitemShut
  {NoStop}%
\bibitem [{\citenamefont {Bondurant}\ and\ \citenamefont
  {Shapiro}(1984)}]{Bondurant1984}%
  \BibitemOpen
  \bibfield  {author} {\bibinfo {author} {\bibfnamefont {R.~S.}\ \bibnamefont
  {Bondurant}}\ and\ \bibinfo {author} {\bibfnamefont {J.~H.}\ \bibnamefont
  {Shapiro}},\ }\href {\doibase 10.1103/PhysRevD.30.2548} {\bibfield  {journal}
  {\bibinfo  {journal} {Phys. Rev. D}\ }\textbf {\bibinfo {volume} {30}},\
  \bibinfo {pages} {2548} (\bibinfo {year} {1984})}\BibitemShut {NoStop}%
\bibitem [{\citenamefont {Sokolov}\ \emph {et~al.}(2001)\citenamefont
  {Sokolov}, \citenamefont {Kolobov}, \citenamefont {Gatti},\ and\
  \citenamefont {Lugiato}}]{Sokolov2001}%
  \BibitemOpen
  \bibfield  {author} {\bibinfo {author} {\bibfnamefont {I.}~\bibnamefont
  {Sokolov}}, \bibinfo {author} {\bibfnamefont {M.}~\bibnamefont {Kolobov}},
  \bibinfo {author} {\bibfnamefont {A.}~\bibnamefont {Gatti}}, \ and\ \bibinfo
  {author} {\bibfnamefont {L.}~\bibnamefont {Lugiato}},\ }\href {\doibase
  https://doi.org/10.1016/S0030-4018(01)01256-1} {\bibfield  {journal}
  {\bibinfo  {journal} {Optics Communications}\ }\textbf {\bibinfo {volume}
  {193}},\ \bibinfo {pages} {175 } (\bibinfo {year} {2001})}\BibitemShut
  {NoStop}%
\bibitem [{\citenamefont {Xiang}\ \emph {et~al.}(2010)\citenamefont {Xiang},
  \citenamefont {Higgins}, \citenamefont {Berry}, \citenamefont {Wiseman},\
  and\ \citenamefont {Pryde}}]{Xiang2010}%
  \BibitemOpen
  \bibfield  {author} {\bibinfo {author} {\bibfnamefont {G.~Y.}\ \bibnamefont
  {Xiang}}, \bibinfo {author} {\bibfnamefont {B.~L.}\ \bibnamefont {Higgins}},
  \bibinfo {author} {\bibfnamefont {D.~W.}\ \bibnamefont {Berry}}, \bibinfo
  {author} {\bibfnamefont {H.~M.}\ \bibnamefont {Wiseman}}, \ and\ \bibinfo
  {author} {\bibfnamefont {G.~J.}\ \bibnamefont {Pryde}},\ }\href
  {http://dx.doi.org/10.1038/nphoton.2010.268} {\bibfield  {journal} {\bibinfo
  {journal} {Nature Photonics}\ }\textbf {\bibinfo {volume} {5}},\ \bibinfo
  {pages} {43 EP } (\bibinfo {year} {2010})}\BibitemShut {NoStop}%
\bibitem [{\citenamefont {Mitchell}(2005)}]{mitchell2005}%
  \BibitemOpen
  \bibfield  {author} {\bibinfo {author} {\bibfnamefont {M.~W.}\ \bibnamefont
  {Mitchell}},\ }\href {\doibase 10.1117/12.621353} {\bibfield  {journal}
  {\bibinfo  {journal} {Proc. SPIE}\ }\textbf {\bibinfo {volume} {5893}}
  (\bibinfo {year} {2005}),\ 10.1117/12.621353}\BibitemShut {NoStop}%
\bibitem [{\citenamefont {Kessler}\ \emph {et~al.}(2014)\citenamefont
  {Kessler}, \citenamefont {K\'om\'ar}, \citenamefont {Bishof}, \citenamefont
  {Jiang}, \citenamefont {S\o{}rensen}, \citenamefont {Ye},\ and\ \citenamefont
  {Lukin}}]{kessler2014}%
  \BibitemOpen
  \bibfield  {author} {\bibinfo {author} {\bibfnamefont {E.~M.}\ \bibnamefont
  {Kessler}}, \bibinfo {author} {\bibfnamefont {P.}~\bibnamefont {K\'om\'ar}},
  \bibinfo {author} {\bibfnamefont {M.}~\bibnamefont {Bishof}}, \bibinfo
  {author} {\bibfnamefont {L.}~\bibnamefont {Jiang}}, \bibinfo {author}
  {\bibfnamefont {A.~S.}\ \bibnamefont {S\o{}rensen}}, \bibinfo {author}
  {\bibfnamefont {J.}~\bibnamefont {Ye}}, \ and\ \bibinfo {author}
  {\bibfnamefont {M.~D.}\ \bibnamefont {Lukin}},\ }\href {\doibase
  10.1103/PhysRevLett.112.190403} {\bibfield  {journal} {\bibinfo  {journal}
  {Phys. Rev. Lett.}\ }\textbf {\bibinfo {volume} {112}},\ \bibinfo {pages}
  {190403} (\bibinfo {year} {2014})}\BibitemShut {NoStop}%
\bibitem [{\citenamefont {Seshadreesan}\ \emph {et~al.}(2015)\citenamefont
  {Seshadreesan}, \citenamefont {Dowling},\ and\ \citenamefont
  {Agarwal}}]{Kaushik2015}%
  \BibitemOpen
  \bibfield  {author} {\bibinfo {author} {\bibfnamefont {K.~P.}\ \bibnamefont
  {Seshadreesan}}, \bibinfo {author} {\bibfnamefont {J.~P.}\ \bibnamefont
  {Dowling}}, \ and\ \bibinfo {author} {\bibfnamefont {G.~S.}\ \bibnamefont
  {Agarwal}},\ }\href {\doibase 10.1088/0031-8949/90/7/074029} {\bibfield
  {journal} {\bibinfo  {journal} {Physica Scripta}\ }\textbf {\bibinfo {volume}
  {7}},\ \bibinfo {pages} {074029} (\bibinfo {year} {2015})}\BibitemShut
  {NoStop}%
\bibitem [{\citenamefont {Lindner}\ and\ \citenamefont
  {Rudolph}(2009)}]{Lindner2009}%
  \BibitemOpen
  \bibfield  {author} {\bibinfo {author} {\bibfnamefont {N.~H.}\ \bibnamefont
  {Lindner}}\ and\ \bibinfo {author} {\bibfnamefont {T.}~\bibnamefont
  {Rudolph}},\ }\href {\doibase 10.1103/PhysRevLett.103.113602} {\bibfield
  {journal} {\bibinfo  {journal} {Phys. Rev. Lett.}\ }\textbf {\bibinfo
  {volume} {103}},\ \bibinfo {pages} {113602} (\bibinfo {year}
  {2009})}\BibitemShut {NoStop}%
\end{thebibliography}
\end{document}